\documentstyle[epsf]{mn}

\newif\ifAMStwofonts

\newcommand{\vektor}[1]{\vec #1}                  
\newcommand{\Msun}{M_{\odot}}
\newcommand{\Lsun}{L_{\odot}}
\newcommand{\Mpc}{\mbox{Mpc}}
\newcommand{\kmpers}{km~s${}^{-1}$}
\newcommand{\hMpc}{{{$h^{-1}$}Mpc}}
\newcommand{\hMpcinv}{$h$Mpc{$^{-1}$}}
\newcommand{\ea}{et\thinspace al.}                        
\newcommand{\cp}{compare}
\newcommand{\eg}{e.g.}                                    

\ifoldfss
  \ifCUPmtlplainloaded \else
    \NewTextAlphabet{textbfit} {cmbxti10} {}
    \NewTextAlphabet{textbfss} {cmssbx10} {}
    \NewMathAlphabet{mathbfit} {cmbxti10} {} 
    \NewMathAlphabet{mathbfss} {cmssbx10} {} 
  \fi
  \ifAMStwofonts
    \ifCUPmtlplainloaded \else
      \NewSymbolFont{upmath} {eurm10}
      \NewSymbolFont{AMSa} {msam10}
      \NewMathSymbol{\upi}     {0}{upmath}{19}
      \NewMathSymbol{\umu}     {0}{upmath}{16}
      \NewMathSymbol{\upartial}{0}{upmath}{40}
      \NewMathSymbol{\leqslant}{3}{AMSa}{36}
      \NewMathSymbol{\geqslant}{3}{AMSa}{3E}

      \let\leq=\leqslant \let\le=\leqslant
       \let\ge=\geqslant
    \fi
  \fi
\fi 

\ifnfssone
  \newmathalphabet{\mathit}
  \addtoversion{normal}{\mathit}{cmr}{m}{it}
  \addtoversion{bold}{\mathit}{cmr}{bx}{it}
  \newcommand{\mbox}[1] {\mathrm{#1}}

  \newmathalphabet{\mathbfit} 
  \addtoversion{normal}{\mathbfit}{cmr}{bx}{it}
  \addtoversion{bold}{\mathbfit}{cmr}{bx}{it}
  \newmathalphabet{\mathbfss} 
  \addtoversion{normal}{\mathbfss}{cmss}{bx}{n}
  \addtoversion{bold}{\mathbfss}{cmss}{bx}{n}
  \ifAMStwofonts
    \ifCUPmtlplainloaded \else
      \UseAMStwoboldmath
      \makeatletter
      \new@mathgroup\upmath@group
      \define@mathgroup\mv@normal\upmath@group{eur}{m}{n}
      \define@mathgroup\mv@bold\upmath@group{eur}{b}{n}
      \edef\UPM{\hexnumber\upmath@group}
      \new@mathgroup\amsa@group
      \define@mathgroup\mv@normal\amsa@group{msa}{m}{n}
      \define@mathgroup\mv@bold\amsa@group{msa}{m}{n}
      \edef\AMSa{\hexnumber\amsa@group}
      \makeatother
      \mathchardef\upi="0\UPM19
      \mathchardef\umu="0\UPM16
      \mathchardef\upartial="0\UPM40
      \mathchardef\leqslant="3\AMSa36
      \mathchardef\geqslant="3\AMSa3E

      \let\leq=\leqslant \let\le=\leqslant
       \let\ge=\geqslant
    \fi
  \fi
\fi 

\ifnfsstwo
  \DeclareMathAlphabet{\mathbfit}{OT1}{cmr}{bx}{it}
  \SetMathAlphabet\mathbfit{bold}{OT1}{cmr}{bx}{it}
  \DeclareMathAlphabet{\mathbfss}{OT1}{cmss}{bx}{n}
  \SetMathAlphabet\mathbfss{bold}{OT1}{cmss}{bx}{n}
  \ifAMStwofonts
    \ifCUPmtlplainloaded \else
      \DeclareSymbolFont{UPM}{U}{eur}{m}{n}
      \SetSymbolFont{UPM}{bold}{U}{eur}{b}{n}
      \DeclareSymbolFont{AMSa}{U}{msa}{m}{n}
      \DeclareMathSymbol{\upi}{0}{UPM}{"19}
      \DeclareMathSymbol{\umu}{0}{UPM}{"16}
      \DeclareMathSymbol{\upartial}{0}{UPM}{"40}
      \DeclareMathSymbol{\leqslant}{3}{AMSa}{"36}
      \DeclareMathSymbol{\geqslant}{3}{AMSa}{"3E}

      \let\leq=\leqslant \let\le=\leqslant
       \let\ge=\geqslant
    \fi
  \fi
\fi 

\ifCUPmtlplainloaded \else
  \ifAMStwofonts \else 
    \def\upi{\pi}
    \def\umu{\mu}
    \def\upartial{\partial}
  \fi
\fi

\title[Large-scale structure formation for power spectra with broken scale
invariance]
      {Large-scale structure formation \\
      for power spectra with broken scale invariance}
\author[R.~Kates, V.~M\"uller, S.~Gottl\"ober, J.P.~M\"ucket, J.~Retzlaff]
       {R.~Kates, V.~M\"uller, S.~Gottl\"ober, J.P.~M\"ucket, J.~Retzlaff
       \\ Astrophysikalisches Institut Potsdam, Germany}
\date{Accepted  Received  ; in original form 1995}
\pubyear{1995}

\pagerange{\pageref{firstpage}--\pageref{lastpage}}

\begin{document}

\maketitle

\label{firstpage}

\begin{abstract}

We have simulated the formation of large-scale structure arising from
COBE-normalized spectra computed by convolving a primordial double-inflation
perturbation spectrum with the CDM transfer function.  Due to the broken scale
invariance ('BSI') characterizing the primordial perturbation spectrum, this
model has less small-scale power than the (COBE-normalized) standard CDM model.
The particle-mesh code (with $512^3$ cells and $256^3$ particles) includes a
model for thermodynamic evolution of baryons in addition to the usual
gravitational dynamics of dark matter.  It provides an estimate of the local
gas
temperature.  In particular, our galaxy-finding procedure seeks peaks in the
distribution of gas that has cooled.  It exploits the fact that ``cold"
particles trace visible matter better than average and thus provides a natural
biasing mechanism.  The basic picture of large-scale structure formation in the
BSI model is the familiar hierarchical clustering scenario.  We obtain particle
in cell statistics, the galaxy correlation function, the cluster abundance and
the cluster-cluster correlation function and statistics for large and small
scale velocity fields.  We also report here on a semi-quantitative study of the
distribution of gas in different temperature ranges.  Based on confrontation
with observations and comparison with standard CDM, we conclude that
the BSI scenario could represent a promising modification of the CDM picture
capable of describing many details of large-scale structure formation.

\end{abstract}

\begin{keywords}
primordial power spectrum -- CDM -- large-scale structure --
galaxy clustering -- cosmological velocity fields
\end{keywords}

\section{Introduction} \label{intr}

The search for a self-consistent model for the formation of large-scale
structure capable of explaining the vast range of available observational data
poses one of the greatest challenges in theoretical astrophysics.  In this
paper, we will present strong evidence that a large body of observations is
consistent with a picture of structure formation based on a
``double-inflationary" model (to be reviewed shortly).  The most directly
relevant observations include the measured large-scale anisotropy of the
microwave background spectrum (Smoot \ea{} 1992, G\'orski \ea{} 1994), power
spectra analysis of galaxy catalogs (Park \ea{} 1992, Vogeley \ea{} 1992,
Fisher
\ea{} 1993), count-in-cell analysis of galaxies (Efstathiou \ea{}, 1990,
Loveday
\ea{} 1992), the galaxy correlation function based on IRAS, CfA, and other
catalogs (Loveday \ea{} 1992, Vogeley \ea{} 1992, Baugh \&{} Efstathiou 1993),
the
angular correlation function of APM galaxies (Maddox \ea{} 1990), abundance of
clusters (Bahcall \&{} Cen, 1993)) as a function of mass, rms line-of-sight
peculiar relative velocity of galaxy pairs (Davis \&{} Peebles, 1983, Mo \ea{}
1993).  Other observational quantities include large-scale velocity fields in
general, the line-of-sight distribution of quasar absorption clouds, X-ray
observations giving information on the temperature and distribution of hot gas,
the observed network of filaments and ``walls", the existence of voids and
their
observed properties, and evidence for the epoch of galaxy and cluster
formation.

Prior to the analysis of the APM catalog and the COBE measurement of the
large-scale microwave anisotropy, the biased, $\sigma_8$-normalized, flat CDM
model (Blumenthal \ea{} 1982) was quite successful in predicting the observed
hierarchical network of filaments, pancakes and nodes and observed galaxy
clustering (White \ea{} 1987).  Biasing, originally devised for explaining the
enhanced correlation of Abell clusters (Kaiser 1984), also had the advantage of
reducing small-scale velocity dispersions to reasonable levels.  These
advantages disappear with the COBE normalization of the CDM spectrum, because
the antibiasing needed to explain the observed variances of counts in cells is
quite difficult to reconcile with our observational knowledge of velocity
fields
and with other tests.  More precisely:  The normalization fixed by the COBE
anisotropy measurement appears to imply excessive power in the small-scale
($<10$\hMpc) regime of the primordial fluctuation spectrum $P(k)$, which
manifests itself in various conflicts with observations.  Predicted local
velocity fluctuations seem to be quite in excess of measured relative pairwise
line-of-sight projected velocities (Davis \&{} Peebles 1983), although the
measurements may be sensitive to sampling effects (Mo \ea{} 1993).

COBE-normalized (flat) CDM simulations also predict much more pronounced and
numerous bound structures such as massive galaxies and galaxy clusters than
observed (Efstathiou \ea{} 1992).  Relative to these massive structures,
COBE-normalized CDM appears to imply fewer galaxies in filamentary structures
than would be expected from observations.  In contrast, as we shall see, the
double-inflationary models reproduce the observed filamentary structures rather
easily and predict fewer massive galaxies and clusters.  The different
evolution
of filamentary structures may be intuitively understood within the Zel'dovich
(1970) picture of anisotropic collapse of overdense regions leading to the
formation of pancakes on a broad range of scales.  In particular, filaments are
transitory structures:  Due to the high normalisation, the COBE-normalized CDM
is further developed, and therefore it is plausible that cooled material in
filaments will have more time to flow into large clumps (\cp{} Doroshkevich
\ea{} 1995).

On the other hand, ``tilted" models (\cp{} \eg{} Cen \ea{} 1993) seem to have
insufficient small-scale power:  as a consequence, the predicted epoch of
galaxy formation is difficult to reconcile with the observation of
high-redshift objects.  Moderate-scale (60\hMpc) power may also be
too low, resulting in rather small streaming velocities.  Mixed dark matter
models (Davis \ea{} 1992, Klypin \ea{} 1993, Cen \&{} Ostriker 1994) appear to
agree better with a large body of observations, but again the late epoch of
galaxy formation may be problematic (Mo \&{} Miralda-Escud\'{e}, 1994,
Kauffmann \&{} Charlot, 1994).

Estimates of hot gas in clusters derived from X-ray observations seem to imply
that a high percentage of the dynamical mass may be in the form of baryons.
E.g., for the Coma cluster, White \ea{} (1994) compute that about 30 \% of the
dynamical mass is in the form of hot gas.  If the estimate for $\Omega_b h^2
\approx 0.013$ obtained from nucleosynthesis is applicable in clusters (unless
significant segregation of hot gas in clusters has occurred), then the total
matter density in the universe would be restricted to the range
$0.1<\Omega<0.3$.  Indeed, low-mass CDM models (Kofman \ea{} 1993, Cen, Gnedin,
\&{} Ostriker 1993) with nonvanishing cosmological constant or low-mass and
open
CDM models (Kamionkowski \ea{} 1994) can explain the enhanced APM angular
correlation function, probably the extended range of positive correlation of
spatial galaxy correlation above 30\hMpc, the cluster mass function, and the
cluster-cluster correlation function; however, the epoch of galaxy and cluster
formation appears to be rather early.  In particular, clusters in open models
tend to be highly developed and tightly bound, exhibiting less substructure
than
observed (Kaiser 1991, Evrard \ea{}, 1993).

In addition to strictly observational constraints, the theoretical appeal of
a model and its consistency with our knowledge of particle physics also play
an important role in its acceptance.  Standard CDM was very appealing because
it involved only one true fit parameter (the amplitude of primordial
fluctuations) together with one phenomenological parameter (biasing).
``Low-mass" models are theoretically less attractive because they require a
nonvanishing cosmological constant in order to achieve a flat universe, as
preferred in an inflationary scenario.  Tilted models introduce one
additional parameter; however, the simplest scenario assuming an exponential
inflaton potential is not theoretically well founded.  Mixed dark matter
models also involve only one additional parameter:  (essentially) the
neutrino mass (since their number density is fixed).  The double-inflationary
model requires two independent parameters.  In Section 2, we review
this model and describe fits to data in the linear regime (Gottl\"ober,
M\"uller \&{} Starobinsky 1991, GMS91; Gottl\"ober, M\"ucket \&{} Starobinsky,
1994, GMS94) and in the Press-Schechter theory (M\"uller 1994a) resulting in
the particular ``BSI" (Broken Scale Invariance) spectrum studied in this
paper.

Encouraged by the success of the BSI model in the linear regime, we now present
the results of numerical n-body simulations in order to test observational
results requiring accurate predictions in the nonlinear regime.  Now, one of
the
subtleties of the nonlinear regime is that our incomplete understanding of
physical processes at relevant length scales causes theoretical uncertainties
that may be comparable to observational uncertainties (cp.  for example,
Ostriker 1993).  Complicated, nonlinear feedback mechanisms govern the dynamics
at scales below our numerical resolution.  In particular, an adequate theory of
galaxy formation (and perhaps evolution) together with significantly improved
computational resources would obviate the necessity for introducing simple
phenomenological concepts such as biasing.  However, an accurate description of
galaxy formation requires a robust treatment of gas dynamics over vast ranges
of
scales, densities, and other conditions; dynamic computation of heating and
cooling, including thermal instability and ionization; star formation,
supernovae; proper handling of the chemical evolution of a multi-component
medium --- to name a few considerations.  Significant progress in applying a
hydrodynamic approach to cosmology has been achieved by Katz \&{} Gunn (1991),
Navarro \&{} Benz (1991), Cen \&{} Ostriker (1992), Katz, Hernquist \&{}
Weinberg (1992), and by Steinmetz \&{} M\"uller (1995).  In particular, a
hydrodynamical description is essential for studying the formation and internal
dynamics of clusters (Evrard 1990).  A comparative review containing additional
references is now available (Kang \ea{} 1994).  An alternative hydrodynamical
approach emphasizing thermal instability and the importance of feedback
mechanisms such as supernovae has been studied by Klypin, Kates \&{} Khokhlov
(1992; KKK92).  Hydrodynamic simulations are clearly superior for obtaining
reliable small-scale predictions from models of large-scale structure
formation.
However, a lack of computer resources still precludes their general use for
high-resolution cosmological applications.

The numerical particle-mesh code used here (Kates \ea{} 1991, KKK91; Klypin
\&{}
Kates 1992, KK) includes a model for thermodynamic evolution of baryons in
addition to the usual gravitational dynamics of dark matter (see Section 3).
The code provides an estimate of the (suitably averaged) local gas temperature
without the enormous computational cost of a full hydrodynamical description.
A
resolution of $512^3$ cells ($256^3$ particles) was achieved.  High resolution
is crucial in the present study in order to include the effects of {\it both\/}
enhanced large-scale power {\it and\/} modest (compared to CDM) small-scale
power in the same simulation.

As discussed in Section \ref{bsi}, considerable preliminary testing using both
linear analysis (\cp{} GMS94) and moderate ($256^3$ cells) resolution (\cp{}
Gottl\"ober 1994, M\"uller 1994b, Amendola \ea{} 1995) was first carried out to
obtain optimal estimates for the parameters $k_{br}$ and $\Delta$ (see Section
\ref{bsi} ).  These preliminary investigations also provided rough estimates of
the sensitivity of our results to ``fine-tuning" of parameters (see
Conclusions), box size, etc.  The power spectrum used in the highest-resolution
simulations and its implementation in the numerical simulations will be given
in
Section \ref{bsi}.

Simulation of the coupling of thermodynamic evolution to gravitational dynamics
provides us with important indirect information concerning observable
quantities; see Section \ref{thermo}.  In Section \ref{cic}, we present our
results on the evolution of the mass distribution and, in Section \ref{temp},
its relationship to thermodynamic evolution.  Of particular interest are the
particle in cell fluctuations, the evolution of clustering with redshift and
its
dependence on the spectrum, redshift-dependence of thermodynamics and
percentage
of cooled particles, the relation between density peaks and hot and cold
particles.

In Sections \ref{gal} and \ref{cluster} we discuss the results of galaxy
identification and galaxy clustering:  galaxy mass function versus
Press-Schechter theory; dependence on spectrum and redshift; galaxy correlation
function versus correlation function of density peaks; clustering and other
properties of galaxy clusters.  The characteristic features of large- and
small-scale velocity fields for BSI spectra are presented in Section \ref{vel}.
In Section \ref{concl}, we draw our conclusions.

\section{BSI Power Spectrum}
\label{bsi}

Standard inflation produces primordial adiabatic density perturbations $P(k)
\propto k^n$ with a spectral index $n = 1$ plus small logarithmic corrections
which stem from the slowly changing horizon length during inflation, i.e.  they
depend on the inflaton potential.  We propose an early cosmological evolution
which produces a power spectrum with a break at a characteristic scale by
introducing more than one effective field responsible for inflation
(Starobinsky, 1985).  We use the combined action of renormalization corrections
and a massive scalar field as the source of a non-flat primordial perturbation
spectrum (GMS91).  This model decouples the clustering properties at large and
small scales.  It requires two new parameters in addition to the amplitude:
First, the ratio of the masses $ \Delta \simeq m/(6.5M) $ appearing in the
Lagrangian

\medskip
\begin{equation}
   L = {1 \over 16 \pi G} (R - {R^2\over 6M^2}) +
    {1 \over 2}(\phi_\alpha \phi^\alpha - m^2 \phi^2) \, ;
\label{Lag}
\end{equation}
\medskip

\noindent
this ratio essentially determines the ratio of power at small (galactic) scales
to that at very large cosmic scales.  Second, the epoch of the transition
between inflationary phases is related to the initial energy density
$m^2 \phi_0^2$ of the scalar field.  This epoch determines a characteristic
length scale $ l_{\mbox{br}} \equiv 2\pi/k_{\mbox{br}}$ where the shape of the
spectrum changes.

In GMS94, linear theory predictions of double-inflationary models were compared
with observational constraints.  Assuming COBE normalization, the tests applied
were as follows:  ``counts in cells" of the IRAS and APM surveys, the APM
galaxy
angular correlation function, bulk-flow peculiar velocities, the Mach number
test, and quasar abundance.  The BSI models were shown to be in good agreement
with all of these tests for the parameter regime defined by $2\le\Delta <4$ and
0.5\hMpc $<k_{\mbox{br}}^{-1}<5$\hMpc, the latter corresponding to length
scales
$l_{\mbox{br}} \approx (3-30)$\hMpc.  Although some freedom is permitted by
these
constraints in the choice of $k_{\mbox{br}}$ (or $l_{\mbox{br}}$), the best fit
seems to be given by $k_{\mbox{br}}^{-1}= 1.5$\hMpc, $\Delta = 3$.  These
values
also led to reasonable effective biasing parameters in preliminary test
simulations.  In all simulation results reported here, the terminology ``BSI
spectrum" refers to these values of the parameters unless otherwise stated.

\begin{figure}
\epsfxsize=234pt \epsfbox{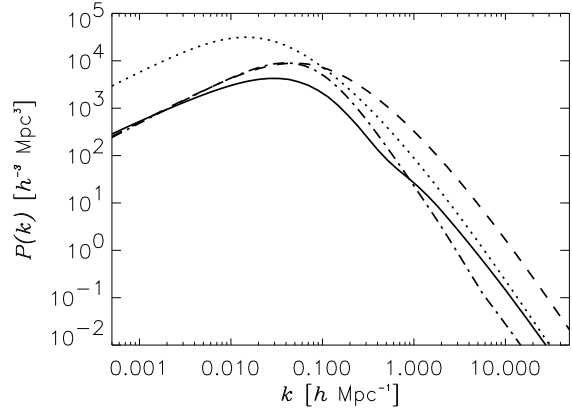}
\caption{BSI power spectrum (solid line) compared with standard CDM (dashed
 line), a $\Lambda$CDM model (dotted line), and a MDM spectrum (dash-dotted
 line).}
\label{fig1}\end{figure}

After convolution with the standard CDM transfer function of Bond \&{}
Efstathiou (1984), taking $\Omega=1$ and $h=0.5$, the power spectrum is used to
generate the initial condition of the numerical simulations described in the
next section.  Fig.~\ref{fig1} plots the transformed BSI spectrum in comparison
with the power spectra of the standard CDM model, a $\Lambda$CDM
($\lambda\equiv
\Lambda/3H^2=0.8$, $\Omega=0.2$) and an MDM model ($\Omega_{\mbox{CDM}}=0.7$,
$\Omega_{\nu}=0.2$, $\Omega_{b}=0.1$).  The standard CDM spectrum has much more
power at all the relevant scales for galaxy formation, while BSI mostly
resembles MDM at large scales, but has more power on small scales, almost as
high as the $\Lambda$CDM model.  An analytical fit to the primordial potential
spectrum is given by

\medskip
\begin{equation}
 k^{3} \Phi (k) = \left\{
       \begin{array}{lc}
       4.2 \times 10^{-6} [\log{(k_s/k)}]^{0.6} + 4.7 \times 10^{-6} &
       (k \leq k_s) \\
       9.4 \times 10^{-8} \log{(k_f/k)} & (k > k_s) \end{array} \right.
\label{Spec}
\end{equation}
\medskip

\noindent
with $k_s = (2 \pi /24)$\hMpcinv{}, $k_f = e^{56}$\hMpcinv{}.  However, for a
proper representation of the transition regime, the numerically tabulated
spectrum is preferable.

\begin{figure}
\epsfxsize=234pt \epsfbox{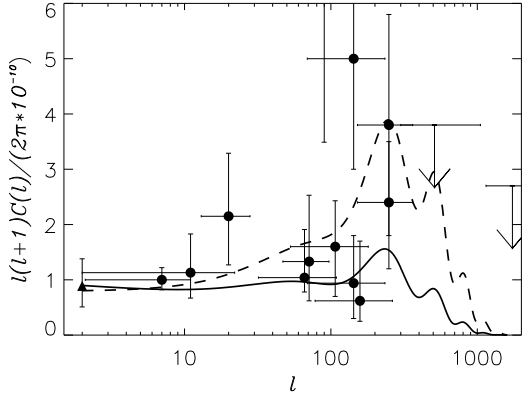}
\caption{Comparison of the cosmic background fluctuation multipoles of BSI
 (solid line) and standard CDM (dashed line) with a series of experiments.}
\label{fig2}\end{figure}

In Fig.~\ref{fig2} we show the predicted multipole moments of the anisotropies
of CMB fluctuations calculated following Gottl\"ober \&{} M\"ucket (1991) in
comparison with new measurements on the angular scale $2 < l < 1200$.  Our
spectra are normalized with the $10^\circ$-variance of the fluctuations
$\sigma_T = (30 \pm 7.5) \,\mu{\mbox{K}}/2.735\,{\mbox{K}}$ from the COBE
first-year data (Smoot \ea{} 1992).  The triangle denotes the first-year COBE
result $Q_{\mbox{rms-PS}} = (16.7 \pm 4)\mu{\mbox{K}}$ (for the spectral index
$n=1$) which is approximately equivalent to our normalization.  The other
experimental data (courtesy of B.~Ratra, from Bond 1994) are from left to right
COBE, FIRS, Tenerife, SP91, SK93, the lower end of the PYTHON error, ARGO, MAX1
and MAX2, full and source free MSAM2 and MSAM3, and the upper limits of WD and
OVRO (for details see Bond 1994).  The reanalysis of the COBE second-year data
by G\'orski \ea{} (1994) implies an about 25 per cent higher normalization of
the spectra (Gottl\"ober 1994).  Consequences of such an increase will be
discussed in the conclusions.

\section{Numerical Realization} \label{pm}

Initial fluctuations corresponding to BSI and standard CDM power spectra were
generated as realizations of a Gaussian random field.  Positions and velocities
were assigned according to the Zel'dovich approximation at redshift $z=25$
(although for larger boxes the Zel'dovich approximation could have been
continued until a later epoch).  The Euler--Poisson system describing the
evolution of self-gravitating, collisionless matter was then evolved from
$z=25$
to $z=0$ using the particle-mesh (PM) code described in KKK91 and KK.

\begin{table*}
\begin{minipage}{120mm}
\caption{Parameters of the simulations}
\begin{tabular}{lrrrrrr}
   \hline
   Simulation & $l_{\mbox{grid}}$ & $M_{\mbox{B+DM}}$ & $\sigma_\delta$ &
   $ \tilde{\sigma}_\delta$ & $\sigma_v$ & $\tilde{\sigma}_v$ \\
   & [\hMpc] & $[\Msun]$ &  &  & [\kmpers{}] & [\kmpers{}] \\
   \hline
BSI-200*& 0.39 & $2.6 \times 10^{11}$ &  2.28  &  2.25  &  490 &  483 \\
BSI-25* & 0.05 & $5.2 \times 10^{08}$ &  4.60  &  4.69  &  223 &  219 \\
BSI-500 & 1.95 & $3.3 \times 10^{13}$ &  0.94  &  0.93  &  512 &  508 \\
CDM-500 & 1.95 & $3.3 \times 10^{13}$ &  3.02  &  2.99  & 1049 & 1040 \\
BSI-200 & 0.78 & $2.1 \times 10^{12}$ &  1.62  &  1.60  &  516 &  479 \\
CDM-200 & 0.78 & $2.1 \times 10^{12}$ &  5.43  &  5.36  & 1103 & 1068 \\
BSI-75  & 0.29 & $1.1 \times 10^{11}$ &  2.59  &  2.54  &  441 &  376 \\
CDM-75  & 0.29 & $1.1 \times 10^{11}$ &  8.79  &  8.65  & 1064 &  979 \\
BSI-25  & 0.10 & $4.1 \times 10^{09}$ &  3.85  &  3.76  &  269 &  219 \\
CDM-25  & 0.10 & $4.1 \times 10^{09}$ & 13.1   & 12.8   &  857 &  735 \\
   \hline
\end{tabular}
\end{minipage}
\end{table*}

To cover a large range of the spectra, we performed a series of simulations
with
different box sizes (see Table 1).  Asterisks denote high resolution
simulations($512^3$ cells and $256^3$ particles); the remainder of the
simulations were performed using $256^3$ cells and $128^3$ particles.  The two
high-resolution BSI simulations were performed in boxes of 200\hMpc{} and
25\hMpc.  BSI-200* is intended for analysis of the large-scale matter
distribution, while BSI-25* should provide good working accuracy on galactic
scales.  In order to highlight differences between BSI and standard CDM, a
series of simulations with half of this resolution were performed using the
same
random seed in boxes of size 25\hMpc, 75\hMpc, 200\hMpc, and 500\hMpc.
$l_{\mbox{grid}}$ gives the size of the particles used in the cloud-in-cell
mass
assignment scheme as well as the formal resolution of the grid on which the
gravitational force is calculated.  The Nyquist wavelength
$2\pi/k_{\mbox{max}}$,
which depends on the {\it number of particles\/} used in the simulation, is
four
times larger than $l_{\mbox{grid}}$.  It corresponds to an upper limit for the
realization of initial perturbations in k-space.  $M_{B+DM}$ is the total mass
of a particle.  Finally, we compare the grid variances of density contrast
$\sigma_{\delta}$ and peculiar velocities $\sigma_v$ of the initial
realization of the Gaussian random field with the expectations of the linear
power spectrum,

\medskip
\begin{equation}
\tilde\sigma_\delta^2 = {1\over2\pi^2}
\int\limits_{k_{\mbox{min}}}^{k_{\mbox{max}}} {\mbox {d}} k\, k^2 P(k),
\quad
\tilde\sigma_v^2 = {H_0^2\over2\pi^2}
\int\limits_{k_{\mbox{min}}}^{k_{\mbox{max}}} {\mbox{d}} k\, P(k),
\label{var}
\end{equation}
\medskip

\noindent
where the numbers are given at $z=0$ transformed according to the linear theory
growth law. They describe the quality of the realizations of the power spectra
in the different simulation boxes.

On the scale of 8 $h^{-1}$Mpc, one often refers to the measured unit variance
of
galaxy counts in spheres as determined from the CfA-catalog ((Davis \&{}
Peebles, 1983).  Using the first integral of Eq.  3 with a top-hat window of
radius 8 $h^{-1}$Mpc{}, we infer $\sigma_{\delta} = 0.46$ for the BSI power
spectrum.  Therefore we predict a (linear) bias of galaxies with respect to
dark
matter of $b \approx 2.2$.  For the standard CDM model we have on the other
hand
$\sigma_{\delta}^2 = 1.12$, i.e.  there the galaxies should be slightly
antibiased.

\begin{figure}
\epsfxsize=234pt \epsfbox{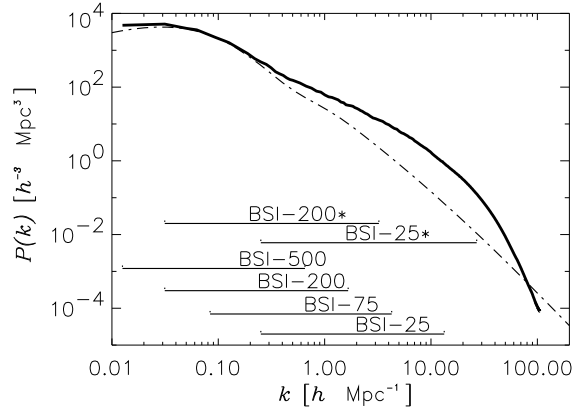}
\caption{Range of spectra realized in the simulations BSI-25 -- 500 (horizontal
lines indicate
${k_{\mbox{min}}}$ to ${k_{\mbox{max}}}$); also shown is the linear theory BSI
spectrum (dash-dotted line) and a reconstruction assembled from the BSI
simulations at $z=0$ (full line).}
\label{fig3}\end{figure}

\section{Thermodynamic estimates of gas temperatures} \label{thermo}

As described in KKK91 and KK, estimates of the average values of the local dark
matter density and velocity and of the local baryon temperature can be obtained
from simulations of large-scale structure (without direct simulation of
hydrodynamics) by following the thermal history of the gas while imagining
that,
smeared out over a sufficiently coarse scale, baryons are transported with the
dark matter; i.e., $\rho_{\mbox{gas}} \equiv \rho \Omega_{\mbox{b}}$, where
$\Omega_{\mbox{b}}=0.1$ is the background fraction of baryons.  The practical
advantage of this picture is that the thermal history of the gas reduces to
integration of one or more ordinary differential equations along {\it known\/}
trajectories (those of the dark matter).

This description is certainly unproblematic before the formation of the first
shocks, when the medium is still cold and fluctuations are small (we put
$T_{\mbox{gas}}=0$ at $z_{\mbox{start}} = 25$).  As perturbations grow,
eventually
the first objects start to collapse, producing caustics in the dark matter and
shocks in the gas.  As simple pancake models show (Shapiro \&{} Struck-Marcell
1985), shocks occur close to caustics.

The label ``shocked'' may be assigned to a particle in one of two ways:  First,
particles for which the Jacobian determinant of the transformation from
Lagrangian to Eulerian coordinates is negative are classified as shocked.
(Particle labels $(i,j,k)$ are the numerical realization of Lagrangian
coordinates.  The position assigned to a particle is the numerical
representation of its Eulerian coordinates.)  The Jacobian is determined for
each particle by considering the volume element constructed from the position
vectors to its nearest Lagrangian neighbors.  Second, a particle inherits the
label ``shocked'' by passing through a shocked region.  More precisely, if
particle $(i,j,k)$ is in cell $(I,J,K)$, then the label shocked is assigned to
particle $(i,j,k)$ if the density of previously shocked gas is nonzero in cell
$(I,J,K)$ and all neighboring cells.

At a shock, the temperature acquired by gas particles with velocity $\vec v$ is
given by
\medskip
\begin{equation}
kT \approx \mu_M m_H (\vektor{v} - \vektor{U})^2/3 ,
\label{tem}
\end{equation}
\medskip
where $\vec U$ is the local velocity determined by interpolation of the
velocity
field onto a coarse grid of twice the usual cell size.  (If it should happen
that too few particles are present to determine $\vec U$, the temperature
assignment is simply postponed.)  This estimate is relatively insensitive to
small errors in the position of the shock.  Here, $m_H$ is the mass of
hydrogen,
and $\mu_M$ is the molecular weight per particle, $\mu_M \approx
(n_H+4n_{He})/(2n_H+3n_{He})\approx 0.6$.

When a particle crosses a shock and is assigned a temperature, we start to
integrate the energy equation along the trajectory of the particle:
\begin{equation}
{{ \mbox{d}T \over \mbox{d}t} = (\gamma -1) \left( {T \over n_H}
{\mbox{d} n_H \over \mbox{d}t} - {\mu_M \over \mu_H}{1 \over k n_H}
(\Lambda _{\mbox{rad}} + \Lambda_{\mbox{Comp}}) \right) },
\label{cool}\end{equation}
where $\mu_H$ is the molecular weight per hydrogen atom, $\mu_H \approx
(n_H+4n_{He})/(n_H)\approx 1.4$, $\rho_{\mbox{gas}} \equiv \rho
\Omega_{\mbox{b}}
=n_H \mu_H m_H$, and $\gamma = 5/3$.  Here, $\Lambda_{\mbox{rad}}$ represents
radiative losses in the hot plasma with assumed primordial abundances, and
$\Lambda_{\mbox{Comp}}$ is the cooling rate due to Compton scattering.  To
estimate $\Lambda_{\mbox{rad}}$, we used analytic fits as in KK and KKK91 to
the
cooling curves given in Fall \&{} Rees (1985).  The medium is treated as
optically
thin and in collisional equilibrium.

Computation of the increment in $T$ due to the (Lagrangian) time derivative of
$n_H$ in Eq.  \ref{cool}{} requires a knowledge of $n_H$ at the present and at
the previous timestep.  For each timestep, we construct the ``gas density'' on
the grid, defined by counting only contributions of shocked (hot) particles.
(For the dark matter density, one of course counts all particles).  For the
evolution of Eq.  \ref{cool}{}, the present value of $n_H$ for Eq.
\ref{cool}{}
is computed simply by interpolation to the position of the particle and then
stored for use at the following timestep.

Now, when material cools below $10^4$~K, stars will form, producing luminous
matter.  By assigning the label ``cooled" to the particles with $T<10^4$~K, we
have a measure for the amount of visible matter.  However, we know that the
efficiency for the conversion of gas to stars is low.  One nonlinear feedback
mechanism limiting star formation could be the energy provided to the medium by
supernova explosions.  A second (also nonlinear) mechanism inhibiting star
formation is ionization and heating due to ambient ultraviolet radiation.  As
emphasized by Efstathiou (1992), a similar mechanism could be responsible for
suppressing the formation of dwarf galaxies.  It should be emphasized that {\it
some\/} mechanism preventing conversion of the gas to luminous matter is
necessary
in BSI models as well as in CDM and in the pre-COBE version of CDM:  otherwise,
all the gas would simply collapse, forming stars and ending up in globular
clusters long before the formation of large objects.

In the present realization of the code, we take these effects into account in a
crude and purely local manner:  we ``reheat" those particles to the temperature
$T_{\mbox{reheat}} = 5\times 10^4$~K which have cooled to below $10^4$~K with a
probability $P_{\mbox{reheat}}$ and assign to the remaining particles the label
``cooled."  Some of the reheated particles later will attain high temperatures,
provided they enter collapsing regions of high density.  Reheated particles can
simply cool again and turn into ``visible" matter.  However, once cooled, a
particle cannot be reshocked. The value $P_{\mbox{reheat}}$ was held constant
at $85$ per cent, the value estimated in KK for a 50 \hMpc{} grid and a
$\sigma_8$ normalized CDM spectrum to give a reasonable fraction of hot gas in
clumps of galactic mass (about 10 per cent).  The fraction of cooled gas
(88 per cent) for a cell size of about 0.05\hMpc{} in the highest resolution
BSI simulation seems to be reasonable.  The distribution of particles of
different temperature ranges, variation with cell size and spectrum, and
related
considerations will be discussed below in Section \ref{temp}.

We recall that the mean density of stars (and other components of the baryonic
matter) in some comoving cell depends in principle on the entire past history
of
the cell and not just on its density at some particular epoch.  As discussed in
KKK91 and KK, comparison of the density distribution of ``cold'' material --
which is related to the density of visible matter -- with the dark-matter
density thus yields information useful for testing cooling as a physical
mechanism for ``biased galaxy formation.''

A reasonable {\it upper\/} bound for the aforementioned ``coarse graining"
(i.e., error in predicting positions of baryons) is the local sound velocity
integrated over a Hubble time (e.g., for $T \approx 5\times 10^6$~K it is 2 or
3\hMpc).  However, in tests -- leaving aside conditions prevailing in clusters
-- for a box of 75 \hMpc{} (cell length 200 $h^{-1}$kpc), the error seems to be
of the order of the cell resolution only.  The explanation seems to be that
both
components move in the same potential well.  The dark matter spends a large
fraction of its time at about the same radius as the gas, because they
originally had the same kinetic energy.  Due to mixing, the dynamics of dark
matter particles bears some resemblance to the dynamics of a gas whose
temperature corresponds to the local velocity dispersion.  Incidentally, from
experiments with truncated spectra in KKK91 and from theoretical
considerations,
we know that temperature predictions are somewhat sensitive to the portion of
the spectrum actually simulated.  This fact implies that the same spectrum
simulated in boxes of varying sizes can yield varying estimates of temperature
distributions.

Results of a 1D-pancake test presented in KKK91 show good agreement of our
model
with hydrodynamical simulations (20 per cent of gas cooled, position of cooling
front and shock wave).  Proper inclusion of hydrodynamics and the effects of
thermal instability would be expected to make a significant difference in at
least two situations:  First, when gas starts to cool efficiently -- this
happens inside dense regions and/or if the temperature becomes too low ($T <
2\times 10^5$~K).  Second, when gas undergoes secondary shocking -- for
example,
in collapse to objects with masses smaller than a galactic mass (for more
detailed comparison, see, KKK92).  Thus, the code does not properly treat the
internal regions of galaxies and clusters, but it would be expected to give a
reasonable approximation for the temperature and distribution of gas which
leaves voids and is trapped in the potential wells of superclusters and
filaments.  Comparison with Cen \&{} Ostriker (1992) supports this expectation:
the baryonic and dark matter distributions look remarkably similar; most of the
gas in regions with density $\rho > 10\langle\rho\rangle$ has a temperature
between $10^6$~K and $10^7$~K, similarly to what was found in KK.  Comparison
with hydrodynamic simulations of periodic disturbances reported in KKK92 also
lead to reasonable qualitative agreement in this regime.

\section{Count-in-cell analysis for dark matter} \label{cic}

One useful method of characterizing the dynamics of clustering is to study the
statistics of counts in cells.  It is a robust measure for distinguishing
different power spectra.  For example, the rms fluctuations of simulated dark
matter particles in a sphere of radius 8\hMpc{} are often used to determine the
linear biasing factor.  The fluctuations on the scale of a galactic halo
provide
an estimate of the threshold density contrast required in galaxy identification
algorithms.  Comparison of the simulated count fluctuations as a function of
radius with rms fluctuations observed in the IRAS catalog (Efstathiou \ea{}
1990a) and the Stromlo-APM survey (Loveday \ea{} 1992) provides an important
test of models.

\begin{figure}
\epsfxsize=234pt \epsfbox{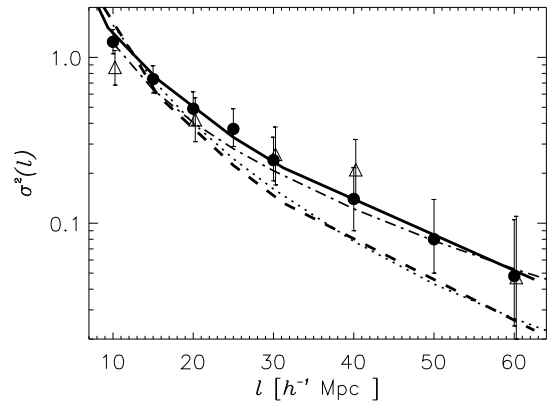}
\caption{Comparison of the variances of counts in cells of galaxies of the
Stromlo--APM (full circles) and IRAS (triangles) survey with the variances of
the dark matter particles in the simulations.  The BSI curve (solid line)
composed from the simulations with different box sizes fits the data quite
well.
The dashed line corresponds to the CDM simulations.  Dashed-dotted and dotted
lines are the linear theory predictions of BSI and CDM, respectively.}
\label{fig4}\end{figure}

In Fig.~\ref{fig4} the curves representing COBE-normalized BSI
have been shifted vertically by a ``biasing" factor $b_{\mbox{BSI}}=2$ in order
to normalize fluctuations to observations at 8\hMpc, while the CDM curves had
$b_{\mbox{CDM}}=0.9$.  CDM fluctuations are too low (significantly outside the
1-$\sigma$ error range on the low side) on scales ($15-50$)\hMpc, after which
the measurement errors increase.  In contrast, the interpolation curves in the
range of BSI simulations fit the data quite well.  The good fit beyond about
15\hMpc{} is consistent with the predictions of linear analysis (GMS94).
The slight differences at small scales are due to the difference between linear
and nonlinear evolution.  The variances of cell counts are calculated in real
space.  At the scales studied the linear theory predicts a constant
amplification factor which can be incorporated in the value of the bias
parameter.  The measured variances provide a good test of the slope of the BSI
power spectrum.  In contrast, the CDM counts in cells are inconsistent with
observations at about the 2-$\sigma$ level.

\begin{figure}
\epsfxsize=234pt \epsfbox{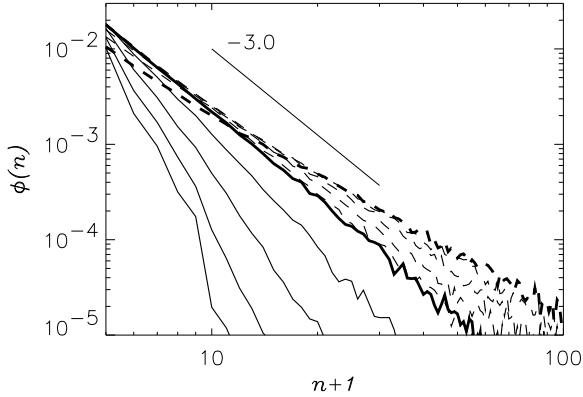}
\caption{Particle-in-cell distribution of BSI-200 (full lines) and CDM-200
(dashed lines) simulations.  The curves ($z=2,1.5,1,0.5,0$) show the dependence
of the
probability distribution of grid cells on the number of particles in the cell;
the curves are steeper at higher redshift.  For reference, a curve of slope
$-3$ is drawn, corresponding to an evolved stage of pancake formation.}
\label{fig5}\end{figure}

The redshift dependence of the probability distribution of particle-in-cell
counts delivers an estimate of the non-linearity of clustering.  Using the
counts on a $128^3$ grid of the 200 \hMpc-simulations, we compare BSI and CDM
simulations in Fig.~\ref{fig5}.  Over four decades in the probability it is
remarkably well described by a power law $ \phi(n) \propto n^{-3}$, which is a
prediction of the pancake model (Kofman \ea{} 1993).  The CDM models at $z=0$
are less steep, corresponding to more big clumps in the matter distribution.
The CDM models at about $z=2$ had a similar slope to the BSI model at $z=0$.
The subsequent evolution to a less rapid fall off is a quantitative measure of
the substantially increased clustering of CDM compared to BSI models.  The
large
dynamical range of the dark-matter probability distribution as computed using
the simulation cells is more of theoretical interest, and it cannot be compared
directly with the one-point distribution function estimated from galaxy
surveys.

\section{Gas temperatures: statistics and spatial structure} \label{temp}

As explained in Section \ref{thermo}, the ``temperature" assigned to each
particle depends on its complete history, including the epoch and conditions
under which it was shocked and the environment (in particular the local
density)
at all intermediate times since shocking.  At a particular time, say $z=0$, the
gas temperature distribution thus contains more information than the density
and
velocity distributions alone.

In interpreting the numerically computed gas temperature distribution, one
should keep in mind the idealizations and limitations involved.  As discussed
above, a proper hydrodynamic treatment of the gas would be expected to make a
systematic difference in regions with secondary shocking and/or very high
density, in particular inside clusters and galaxies.  However, any treatment of
the gas, even including hydrodynamics, will contain numerical inaccuracies
(resulting both from limited resolution and from systematic errors) as well as
theoretical uncertainties due to incomplete modelling.  Bearing all of these
errors and uncertainties in mind, we take a brief look at what can be learned
from particle temperature statistics.

\begin{figure}
\epsfxsize=234pt \epsfbox{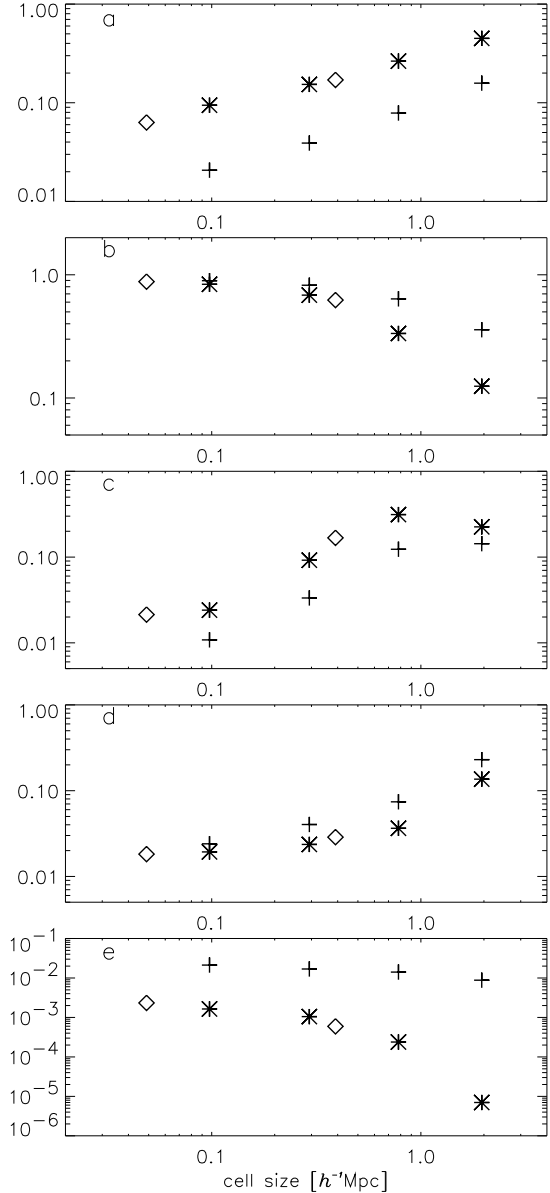 }
\vskip 0.5cm
\caption{Percentage of particles in different temperature ranges as a function
of {\it cell\/} size at redshift $z=0$.  Plus signs indicate CDM simulations,
asterisks BSI, and diamonds high-resolution (512$^3$ cells) BSI simulations.
a) never shocked, b) cooled, c) temperature range $10^4 < T < 5 \times10^4$,
d) temperature range $10^5 < T < 1.5 \times10^6$, e) temperature range
$T>10^7$.}
\label{fig6}\end{figure}

In Fig.  \ref{fig6}, the percentage of particles in selected temperature ranges
at $z=0$ for BSI and CDM simulations is plotted against cell size, which for
all
simulations shown is inversely proportional to the highest resolved wavenumber.
Indeed, studies of truncated spectra in high-resolution simulations (KKK91)
indicate that the wavenumber limit is crucial to temperature statistics, i.e.,
qualitatively similar trends would be observed even if the spatial resolution
were to be improved while keeping the wavenumber limit constant.
(Incidentally,
percentages do not add up to one, because the temperature ranges shown are not
exhaustive.)  Plus signs indicate CDM simulations, asterisks and diamonds BSI;
diamonds indicate the high-resolution simulations, in which the ratio of
highest
to lowest resolved wavenumber is twice as large.  One expects the (limited)
spectral range to have an influence on gas temperature statistics, and this
expectation is borne out in Fig.  \ref{fig6}.  The trend with respect to cell
size is different in different temperature ranges.  However, the most striking
regularity is that in all temperature regimes, a smooth (in most cases
monotonic) trend is apparent upon plotting percentage against {\it cell\/} size
(equivalently here:  {\it highest\/} resolved wavenumber) rather than {\it
box\/} size (equivalently here:  {\it lowest\/} resolved wavenumber).  A plot
with respect to box size would be obtained by shifting the high-resolution
points (diamonds) by one point to the right, which in all cases would move them
off an otherwise smooth curve.

\begin{figure}
\epsfxsize=234pt \epsfbox{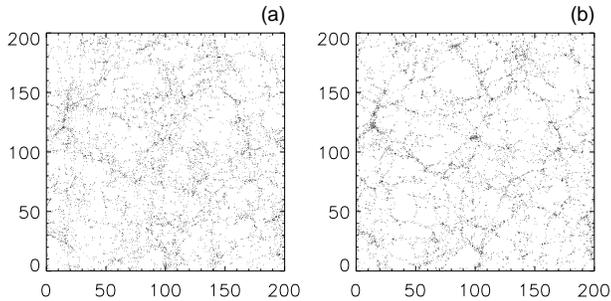}
\caption{Comparison of slices through 200\hMpc{} BSI (a) and CDM (b)
simulations
in the temperature range $10^4$~K $< T < 5\times 10^4$~K.}
\label{fig7}\end{figure}

\begin{figure}
\epsfxsize=234pt \epsfbox{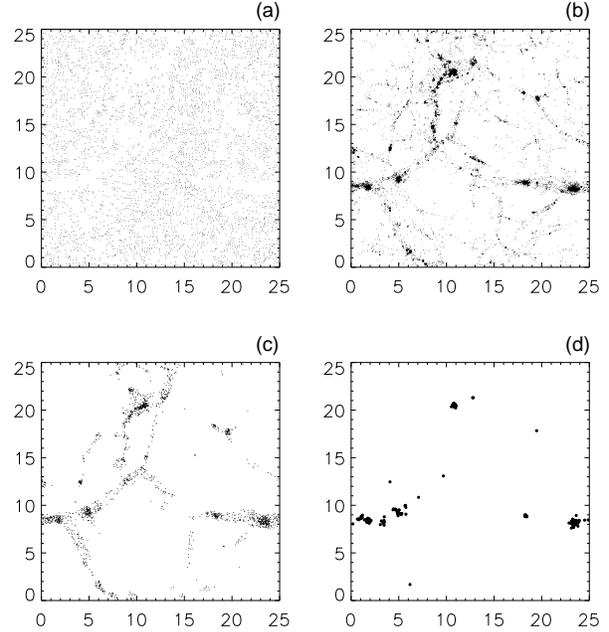}
\caption{Slices of BSI-25* simulation at z=0.  a) unshocked; b) cooled;
c) $10^5$~K $< T < 1.5\times 10^6$~K; d) $T>10^7$~K.}
\label{fig8}\end{figure}

Fig.  \ref{fig7} illustrates a typical slice in the range $10^4$~K $< T < 5
\times10^4$~K for both BSI and CDM simulations.  (Note that the random seed
used
in BSI and CDM simulations was the same, resulting in the same phase relations
for each perturbation mode.)  Slices are about 12\hMpc{} thick.  About 9000
particles are plotted in each slice (all of the particles in the CDM slice and
about 40 per cent of the particles in the BSI slice).  In Fig.  \ref{fig8},
spatial distributions for BSI simulations are illustrated in the same thin
slice
for particles in the ranges $10^5$~K $< T < 1.5 \times10^6$~K, $T>10^7$~K,
cooled ($T < 10^4$~K), and never shocked.  The different temperature ranges
highlight quite different features of the distribution in a way that would not
be possible using simple peak statistics.

\noindent We now consider various temperature ranges in detail:

\noindent {\bf Never shocked:}
This percentage is a measure of the fraction of particles not in pancakes.  The
fraction increases monotonically with cell size for both BSI and CDM (Fig.
\ref{fig6}a).  The trend may be attributable to the fact that for larger cell
size, the portion of the spectrum actually simulated is shifted toward smaller
wavenumbers, with correspondingly smaller values of $k^3 P(k)$, resulting in
longer timescales for formation of structures such as pancakes.  The number of
unshocked particles is consistently lower by a factor of about three for CDM
than for BSI, directly reflecting the significantly enhanced small and
medium-scale power of CDM compared to BSI.  In Fig.  \ref{fig8}a, the spatial
distribution of unshocked particles is shown for a 25\hMpc{} high-resolution
simulation.  The distribution shows practically no structure, aside from being
absent in regions of high density, where all of the particles are shocked.
Visual inspection of slices confirms one's expectation that material in voids
consists mainly of unshocked particles.

\noindent {\bf Cooled:}
{}From plausibility arguments one expects the distribution of ``cooled"
particles
to trace the galaxy (halo) distribution more accurately than does the unbiased
dark-matter distribution.  (A higher probability of cold gas obviously favors
star formation.)  The results of KKK91 in the context of high-resolution, 2D
simulations lend additional support to this expectation:  There, halos were
found with high confidence by a (``friends-to-friends") clustering algorithm
applied to {\it all\/} particles.  The percentage of cold particles inside
halos
was significantly higher than outside.  Moreover, the epoch of galaxy formation
(estimated by cluster analysis at various redshifts) agreed with the average
redshift $z_{\mbox{gal}}$ at which the cold particles in galaxies ``cooled."
In
three dimensions, we exploit the affinity of cooled particles for galaxies in
our galaxy-finding routines:  As discussed in Section \ref{gal}, the routines
begin by searching for peaks in the cooled-particle distribution.

As seen in Fig.  \ref{fig6}b, the percentage of cooled particles at $z=0$
varies
strongly with cell length (more precisely, with Nyquist frequency), mainly
because power at smaller length scales leads to early formation of small
pancakes and, in spite of ``reheating," subsequent cooling.  (The redshift at
cooling is recorded for each cooled particle.)  Consider a structure of
characteristic size $L$.  Smaller structures are associated on the average with
smaller characteristic initial temperatures ($\propto L^2$, since typical
relative velocities in Eq.~\ref{tem} scale with $H_0 L$).  The CDM simulations
have more small-scale power and therefore consistently fewer unshocked
particles
and more cooled particles.  In KKK91, it was shown that truncation of the
spectrum below the Nyquist frequency leads to significant reduction of the
number of cooled particles (keeping spatial resolution constant).  Thus, as in
the case of unshocked particles, the variations with ``cell size" seen here can
probably be attributed to changes in the range of the spectrum actually
simulated.  Filaments and voids are evident in the cold particle distribution;
the hierarchy of filament separations and void sizes give the visual impression
of a broad distribution of characteristic length scales.

\noindent{\bf Particles in the temperature range $(1-5)\times 10^4$~K:}
This temperature range would tend to be associated with Lyman-$\alpha$ clouds.
However, a more realistic treatment of the effect of the ultraviolet background
on the matter (including ionization and heating) would be required to draw
quantitative conclusions concerning Lyman-$\alpha$ clouds.  Such a treatment is
discussed by (M\"ucket \ea{} 1995).  Nevertheless, we may obtain some hints as
to what to expect from a more realistic treatment by examining the trend in
Figs.  \ref{fig6}c.  The fraction grows monotonically both in BSI and in CDM
with cell scale, except for the 500\hMpc BSI simulation, with CDM consistently
lower.  A reasonable interpretation may be that earlier pancake formation and
efficient cooling at moderate redshifts in the present code deplete the
reservoir of particles (especially those near galaxies) that would otherwise be
candidates for this temperature range.  The turnover of the BSI curve at large
scales is not surprising, since at the largest scales, the reservoir of shocked
particles sets the limit.  (In any case, quantitative studies of Lyman-$\alpha$
clouds would require resolution of about $50-100$ kpc.)

With proper treatment of ionization and heating, the relative fraction of cold
particles would be expected to drop significantly in favor of particles in the
range of temperatures $(1-5)\times 10^4$~K, except in very dense regions.  Upon
examination of numerous slices, there is the visual impression that particles
in
the temperature range $(1-5)\times 10^4$~K at least roughly trace the
distribution of cold particles at $z=0$.  This may be a hint that a population
of Lyman-$\alpha$ clouds could be associated with galaxies (cp.  Petitjean
\ea{}
1995).

Comparing BSI and CDM slices for the temperature range considered in Fig.~7,
there is a strong impression of thinner, better defined filaments and emptier
voids in CDM than in BSI simulations.  (Note that the total number of particles
is the same.)  This impression is consistent with the interpretation that the
standard CDM looks like the BSI model would look like if it were evolved to a
larger fluctuation amplitude.

\noindent{\bf Particles in the temperature range $10^5$~K $<T< 1.5 \times
10^6$~K:}
As explained in KK, there are strong theoretical and observational reasons for
an association between gas in this ``warm" temperature range and filamentary
structures.  In Fig.  \ref{fig8}c, it is evident that warm particles trace
filaments quite closely in BSI.  The same qualitative picture is found in all
boxes.  Comparison of CDM and BSI simulations in this warm regime shows little
qualitative difference in filamentary structures, except that CDM consistently
leads to somewhat more gas in the warm regime than BSI.  This occurs despite
CDM's enhanced accumulation of material in clusters, which would tend to
transfer gas from the warm to the hot regime (see discussion of hot gas below).
The trend toward more warm gas at larger simulations may be related both to the
increased reservoir of shocked, but not yet cooled particles (an essentially
small-scale effect) and perhaps the influence of large-scale power in the
portion of the spectrum actually simulated.  This second cause is consistent
with a rise just visible in the BSI curve at cell-size 0.39\hMpc{}
(high-resolution simulation).

\noindent{\bf Particles in the temperature range $ T > 10^7$~K:}
The ``hot" gas represented by these particles is generally associated with the
deep potential wells of clusters, and therefore it is no surprise that in Fig.
8d hot gas traces clusters (also verified by comparison with galaxy catalogs).
The percentage of ``hot" particles with $T>10^7$~K, which are nearly all
associated with clusters or large galaxy groups, is consistently an order of
magnitude higher for CDM than for BSI, the trend becoming even stronger at
larger cell size.  This trend is consistent with the general conclusion that
CDM
causes more and larger massive structures (also evident in slices).  Moreover,
these structures seem to be farther evolved in CDM, so that deep potential
wells
would also tend to be broader.  Broader cluster potential wells would explain
why the disparity between CDM and BSI grows at larger cell lengths (Fig.
\ref{fig6}e).  Cluster potential wells are still fairly well resolved at the
poorest resolution in CDM, but not in BSI.

\medskip

\section{Galaxy Identification and Statistics} \label{gal}

The galaxy finding procedure used here was a modified density peak
prescription:
Based on the results of KKK91, KK, cooled particles are preferred tracers of
the
galaxy distribution.  We therefore began by searching for local maxima of the
density field of cold particles exceeding a predetermined threshold value
$\delta_{\rm th}$.  (The local maxima were also required to be maxima within a
5$^3$ cell neighborhood.)  The mass of the corresponding galaxy was then
computed by summing the contributions of {\it all\/} particles within 0.5 cell
lengths of the maximum in the lower resolution simulations and within one cell
lengths of the maximum in the high resolution simulations.  This leads to the
same physical length for selecting halos in simulations of the same box size,
i.e. to comparable catalogues.  Since the centroid can shift due to addition of
new particles, the procedure was iterated six times, and a centroid, velocity,
and mass were assigned to the galaxy.  For each simulation, several galaxy
catalogs were constructed using various choices of threshold.  Appropriate
choices lie in the range $\delta_{\rm th}=(1.5-3)\sigma_\delta$.  In the end,
one seeks a catalog including at least as many galaxies as would be expected
from
observational counts of galaxies with luminosities exceeding the characteristic
Schechter luminosity $L_*$ in the given volume.  For numerical simulations, the
mass corresponding to ``$L_*$"-galaxies can only be determined from this
requirement {\it a posteriori}.  We observed that reducing the threshold simply
resulted in including additional galaxies at the low end of the mass spectrum,
but had virtually no effect on the galaxies at higher masses.  Thus we can
extract one big galaxy catalog from each simulation, which can be cut off at
the
low-mass end as required.  A natural cut-off is implied by the finite
resolution
of the simulations.  At later stages, this cut-off grows due to the overmerging
effect.

\begin{figure}
\epsfxsize=234pt \epsfbox{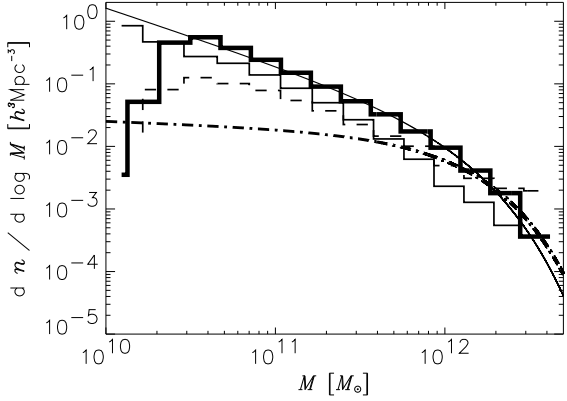}
\caption{The 'galaxy' mass function of BSI-25 (thin solid histogram), of
BSI-25*
(thick solid histogram) compared with the CDM-25 function (dashed histogram).
The solid and dash-dotted curves correspond to comparison Schechter curves
with parameters provided in the text.}
\label{fig9}
\end{figure}

In Fig. {\ref{fig9}} we compare mass functions derived from the small box
simulations where
we expect the most reliable identification of single galactic halos. The halo
masses span the range from $10^{10} \Msun$ to $4 \times 10^{12} \Msun$, and
they
can be fitted by Schechter curves of the form
\medskip
\begin{equation}
\mbox{d} n / \mbox{d} \log M = n_* (M/M_*)^{-p} \exp (-M/M_*) ,
\label{Schechter}
\end{equation}
\medskip
\noindent For BSI-25* we get $M_*=10^{12} \Msun$, $n_*=2.6 \times 10^{-2}
h^3\Mpc^{-3}$ and $p=0.8$, \cp{} the solid line in Fig. \ref{fig9}.  The lower
dashed line represents the observational results from Efstathiou \ea{} (1988),
$n_*=1.6 \times 10^{-2} h^3\Mpc^{-3}$, $p=0.1$, and also $M_*=10^{12} \Msun$
(assuming a quite high $M/L$ ratio of about $100 \Msun/\Lsun$).  Similar
parameters of the Schechter function are derived in Loveday \ea{} (1992).

\medskip
\begin{figure}
\epsfxsize=234pt \epsfbox{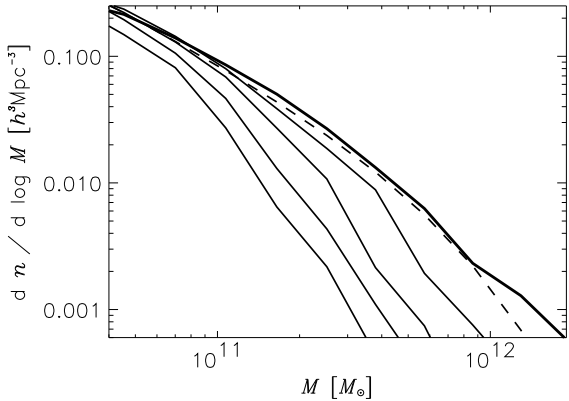}
\caption{The halo mass function of BSI-25 at redshifts $z=2,1.5,1,0.5,0$.  For
$z=0$ a best fit Schechter distribution is shown using a characteristic mass
$M_*=6 \times 10^{11} \Msun$, a power index $p=-1.1$ and mass density
$n_*=1.4\times 10^{-2} h^3\mbox{Mpc}^{-3}$.}
\label{fig10}\end{figure}

The simulated mass spectra lead to approximately the same abundance of
$M_*$ galaxies as the observations.  Comparison of galaxy mass spectra between
CDM and BSI simulations {\it at the same resolution\/} shows that the CDM
leads to a slower inclination at small masses, despite of the excessive power
at small scales.  Due to overmerging (Katz \&{} White 1993; Kauffmann \ea{}
1993), however, we know that spectra from PM simulations may not be directly
compared with observational data.  One way to imagine how overmerging comes
about is as follows:  Potential wells associated with {\it galaxies\/} can
generally bind particles with velocities of about 250 \kmpers.  However, the
potential wells in clusters can bind particles with velocities of about 1000
\kmpers.  The probability that a particle within a cluster will be near one of
the deeper troughs may be slightly elevated, but not enough to distinguish
galaxies, at least not using the present scheme at the presently attainable
resolution.  In a hydrodynamic simulation with proper treatment of cooling
(Katz
\&{} Weinberg, 1992, Katz \&{} White, 1993, and Evrard, Summers \&{} Davis
1994), one would expect baryons to condense earlier in the galaxy (halo)
potential wells, before the galaxies are assembled in the cluster.  On the
other
hand, the PM dynamics leads to an merging of halos per se.  We do not intend to
make predictions on the galaxy mass functions in BSI from our simulations.  We
only use the halo identification scheme to get a catalog of objects with a
reasonable number density, whose clustering properties should be characteristic
for the primordial perturbation spectrum.

\begin{figure} \epsfxsize=234pt \epsfbox{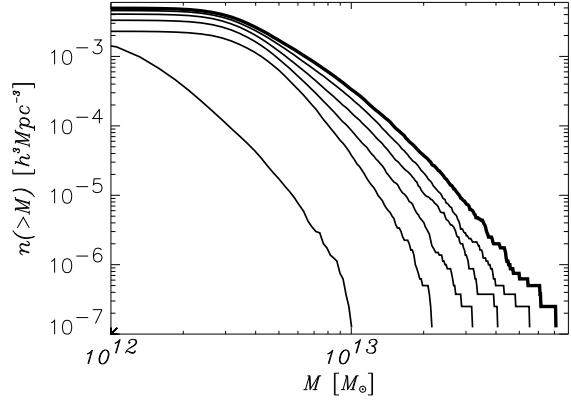}
\caption{The cumulative halo mass function of BSI-200* at redshifts $z=1.8,
0.67, 0.43, 0.25, 0.1, 0$.  The mass range corresponds to small groups of
galaxies which undergo substantial growth after $z=1.8$.}
\label{fig11} \end{figure}

In order to understand the development of (over-) merging in a qualitative way,
we studied the evolution of the mass spectra of halos (massive galaxies and
groups).  The redshift dependence of the differential galaxy mass function of
BSI-25 is shown in Fig.  \ref{fig10}.  A Schechter fit is possible for all
redshifts, only at $z=0$ the number of very massive halos begins to exceed the
exponential fall-off.  Note that while the number of low-mass halos ($M <
10^{11} \Msun$) grows by less than a factor of two between $z=1$ and $z=0$, the
number of galaxies of $M>5 \times 10^{11} \Msun$ grows by a factor of about 10
or more.  The redshift dependence of cumulative mass functions for BSI-200* is
shown in Fig. \ref{fig11}.  Focusing our attention on the behavior of the
curves
beyond $2\times 10^{13}\Msun$, we found an increasing of the number of
high-mass
halos after about $z=0.67$, suggesting  that at earlier epochs (before cluster
formation), overmerging may be less severe.

Summarizing, CDM appears to lead to more massive objects than BSI (\cp{} in
Fig.
{\ref{fig9}} the thin solid and dashed lines for BSI and CDM distributions,
respectively). However, in view of overmerging, the apparent disagreement of
mass
spectra derived in both scenarios with mass spectra inferred from luminosity
functions
cannot be interpreted as a weakness of either model at the present stage.

We can gain important information from the spatial distribution and velocities
of galaxies that are derived from our catalogs.

\begin{figure}
\epsfxsize=234pt \epsfbox{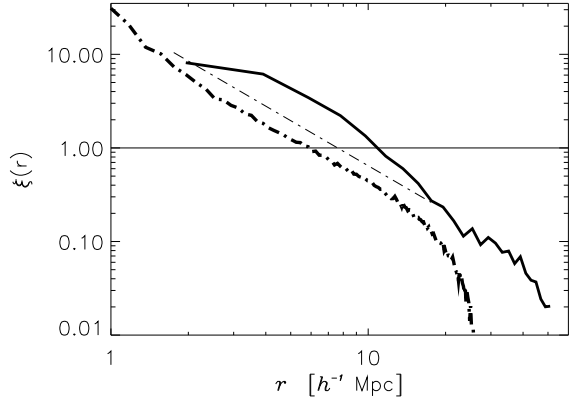}
\caption{The correlation function of galactic halos (identified in
BSI-200*, (dash-dotted line) and of cluster halos (using BSI-500, solid line).
For comparison we show a straight line with slope 1.6}
\label{fig12}\end{figure}

In Fig.  \ref{fig12}, we show the two-point correlation function for the galaxy
catalog constructed from the simulation BSI-200*.  It is well described by a
power law $\xi=(r/r_0)^{-\gamma}$ for 1\hMpc{} $< r <$ 15\hMpc, with slope
$\gamma\approx 1.6$ and correlation length $r_0 \approx 6$\hMpc.  Both values
are in reasonable agreement with galaxy surveys.  (Because the existence of a
power
law for the slope of the correlation function is a quite stable property of
hierarchical clustering, as demonstrated in numerous simulations, it is not a
strong discriminator between models.)  The slope is a bit on the shallow side,
indicating that our spectrum could have a slight power deficit at galaxy
scales.  The correlation radius is also bit high; the halos in this galaxy
catalog may be biased toward high masses and probably include small groups.
Nevertheless, considering the uncertainties in identifying galaxies and the
effect of overmerging, the BSI simulation results are quite compatible with
observations.  On the contrary, the correlation function for 'galaxies'
identified in the COBE-normalized CDM simulations is much too steep (and one
might expect that without overmerging it would have been even steeper).
Finally, the break in the correlation function at 20\hMpc{} and the zero
crossing over 25\hMpc{} are quite remarkable.

\begin{figure}
\epsfxsize=234pt \epsfbox{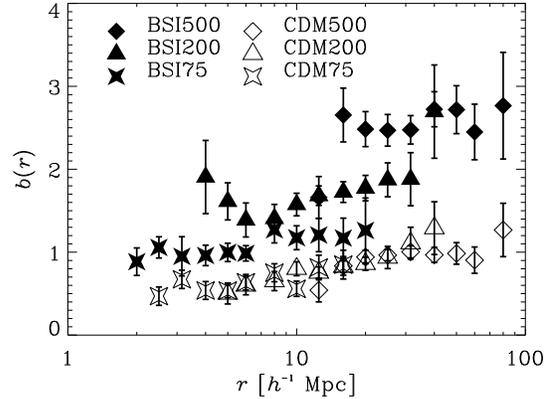}
\caption{Integral bias of different BSI vs.  CDM simulations, the error bar are
$1-\sigma$ errors estimated from the count in cell variances.}
\label{fig13}\end{figure}

A robust measure of clustering is provided by the count-in-cell variances,
which
are volume averages of the two-point correlation function.  The (mass-weighted)
variance of galaxy counts in spheres of radius 8$h^{-1}$Mpc{} in the BSI-200
simulation is consistent with the observed variance 1.  We compare the
variances
of galaxy and dark matter counts for spheres of different radii to get a
measure
of the integral bias,
\begin{equation}
b(r) = \sigma_{\mbox{gal}}(r)/\sigma_{\mbox{DM}}(r).
\label{bias}\end{equation}
To minimize the effect of overmerging, we use in any case mass-weighted cell
variances.  Fig.  13 shows this bias parameter as a function of the chosen
length scale.  For the CDM models with more power on galactic scales, we obtain
a slight antibiasing, while the bias of the BSI halos varies from 1 to 2.5, if
one inspects the simulations BSI-25 to BSI-500.  It should be noted that the
bias factors within one simulation are almost constant.  For deriving the
counts-in-cells, we used randomly placed spheres with a total volume not
exceeding the sample volume.  Therefore we obtain statistically significant
results.  The mean errors in the cell counts are used to estimate the error
bars
in the 'bias' in Fig.  13.  The somewhat surprising antibias of the
'peak-selected' galaxies in CDM is connected with the low threshold of the
galaxy catalog used in this analysis (compare the mass functions of the
galaxies).  On the other hand this result is in accordance with the
normalization of the corresponding simulations, the comparison of the linear
theory $\sigma_{DM}(8 h^{-1} \mbox{Mpc})$ variances, cp.  Fig.  4, and the
measured unit variance of galaxy counts.  The results of the integral bias are
summarized in Table 2.

\begin{table*}
\begin{minipage}{120mm}
\caption{Integral bias of galaxy halos}
\begin{tabular}{lccc}
   \hline
   Simulation &  radii of spheres $r$ & mean bias & standard \\
   & \hMpc & $ \langle b \rangle$ & deviation \\
   \hline
BSI-200* & ~5~~ -- ~40~~  & 1.4 & 0.2 \\
BSI-500  & 12.5 -- 100~~  & 2.5 & 0.5 \\
BSI-200  & ~5~~ -- ~40~~  & 1.7 & 0.4 \\
BSI-75   & ~2~~ -- ~12.5  & 1.0 & 0.1 \\
BSI-25   & ~0.8 -- ~~5~~  & 0.8 & 0.3 \\
CDM-500  & 12.5 -- 100~~  & 0.9 & 0.3 \\
CDM-200  & ~5~~ -- ~40~~  & 0.8 & 0.2 \\
CDM-75   & ~2~~ -- ~12.5  & 0.6 & 0.1 \\
CDM-25   & ~0.8 -- ~~5~~  & 0.5 & 0.2 \\
   \hline
\end{tabular}
\end{minipage}
\end{table*}

\section{Analysis of Galaxy Clusters} \label{cluster}

\begin{figure} \epsfxsize=234pt
\epsfbox{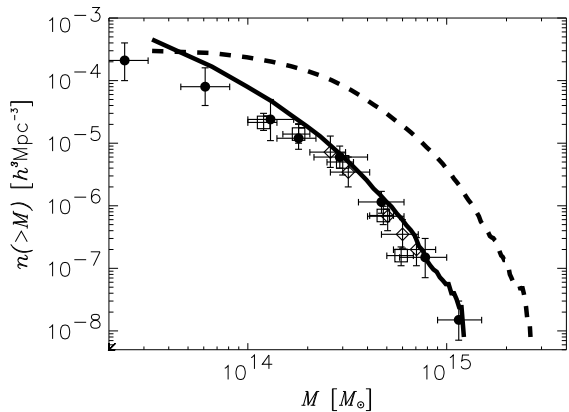} \caption{The cumulative number density of cluster halos of
the BSI model (solid line, BSI-500) is a good description of the data of Cen
\&{}
Bahcall (1993); full dots:  optical data mainly from Abell clusters; squares:
X-ray observations from EXOSAT and HEAO II.  CDM simulations (dashed line)
significantly overproduce cluster halos.}
\label{fig14} \end{figure}

Clusters of galaxies are widely acknowledged as a sensitive test of
cosmological
scenarios.  Their formation is connected with the very recent decoupling of
large masses from cosmic evolution.  Henry and Arnaud (1991) were the first to
derive restrictions on the power index of the primordial perturbation spectrum
from X-ray cluster data.  Using a catalog of 25 clusters with X-ray flux $L_X
\ge 3 \times 10^{-11}$ erg cm$^{-2}$s$^{-1}$ (flux measured in the range $2 -
10$ keV), they derived the dependence of the spatial number density on the
X-ray
temperature.  Since the latter can be used as an indicator of of the cluster
mass (cp.\ the estimates of Evrard 1990), they obtained an estimate of the
condensation probability of different mass scales.  The results of Henry and
Arnaud (1991) are based on the Press--Schechter (1974) theory, which supposes a
Gaussian distribution of rare mass fluctuations when smoothed with a top-hat
filter of the respective scale.  Henry \&{} Arnaud used in their analysis a
power
law spectrum $P(k) \propto k^n$ from which they obtained a spectral index $n
\simeq -1.7$ ($-2.4 < n < -1.3$) at $k \simeq 0.3h $ \Mpc$^{-1}$.  This is
shallower than the standard CDM model on the relevant scale (\cp{} also
Bartlett
\&{} Silk, 1993).  Here we identify galaxy clusters in the largest simulations
(BSI/CDM-500) using a effective search radius of 1.5\hMpc, the Abell radius.
The resulting mass function of cluster halos is shown in Fig. \ref{fig14}.  The
observational points stem from a collection of both optical and X-ray data
superimposed to get an universal empirical mass function (Cen \&{} Bahcall
1993).
The BSI-simulation fits the observational data quite well, although at $M < 6
\times 10^{13} \Msun$ we see a slight overproduction of clusters.  Cluster
formation is a quite recent process, as Fig.  \ref{fig11} demonstrates.  The
CDM model
leads to a tremendous overproduction of halos; in particular, the mass function
is too steep on the cluster scale, this conclusion is typical for $\Omega=1$
models.  The break length of BSI is strongly restricted by the cluster mass
function as shown by a Press--Schechter analysis (M\"uller 1994a).

The cluster number density is a sensitive test of the power spectrum at the
transition scale between the different inflationary stages.  An independent
test
of the power spectrum on these scales is the cluster-cluster correlation
function (Bahcall \&{} Soneira 1983, Klypin \&{} Kopylov 1983).  While the BSI
correlation function, shown in Fig. \ref{fig12} as a dashed line, has the
correct slope, it has a comparatively small correlation radius, $r_0 \approx
10$
\hMpc.  Only slightly larger correlation radii are derived by Dalton \ea{}
using
the APM survey ($r_0=14.3 \pm 2.35 $ \hMpc) and by Collins \ea{} (1994) for a
ROSAT selected cluster survey.  Our estimate of $r_0 \approx 10 $\hMpc{} may be
marginal consistent with the optical catalog.  Of similar importance is the
range of the region of positive correlation of cluster-cluster correlation.  We
find from our simulation a positive cluster-cluster correlation up to at least
80\hMpc.  The same seems to be indicated by (up till now somewhat uncertain)
observational results.

\section{Large and small-scale velocity fields}
\label{vel}

\begin{figure}
\epsfxsize=234pt \epsfbox{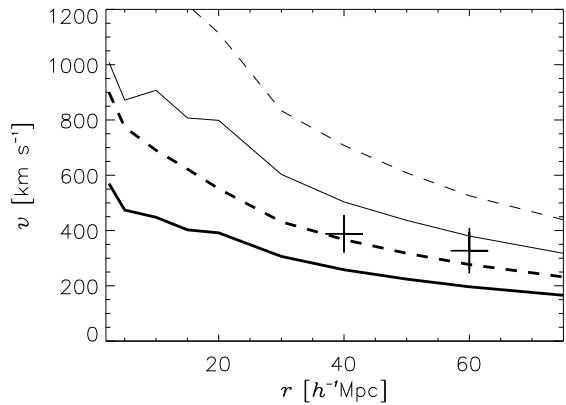}
\caption{Streaming motions for BSI--500 (lower solid line) and CDM--500 (lower
dashed line) for the largest simulation box.  The thin solid line and dashed
line
delimit the {\it upper\/} one-sigma range obtained from different observers in
the BSI--500 and CDM-500 simulation, respectively.  The two error bars apply to
the data from Bertschinger \ea{} 1990).}
\label{fig15}\end{figure}

Using catalogs of objects (large galaxies or clusters) constructed from the
200\hMpc{} ($M>2 \times 10^{12} \Msun$) and 500\hMpc{} ($M>3 \times 10^{13}
\Msun$)
simulations, we estimated typical bulk velocities (and standard deviations) by
computing unweighted average velocities in spheres of radius $R$ (from 2.5 to
75\hMpc) centered about 512 equally spaced observers.  A top-hat filter was
used.
(Note that with $8^3$ spheres considered there is of course some overlap in the
larger spheres.) The results of the 500 \hMpc{} simulation are shown in
Fig.~\ref{fig15}. Bertschinger \ea{} (1990) estimated
the average velocities of the mass within spheres of radius 40 and 60\hMpc{}
centered at the Local Group as $388 \pm 67$ and $327 \pm 82$ \kmpers,
respectively (see data points in Fig.~\ref{fig14}).  At the 1-$\sigma$ level,
these estimates are compatible both with the BSI results and with CDM.  The rms
velocities for BSI-200* (not plotted) and BSI-500 are both well fitted by
curves
of the form $ v_{\rm rms}=v_0 \exp(-x/x_0)$, with $v_0 \approx 490$ \kmpers,
$x_0=41.7$\hMpc{} for simulation BSI-200*, and $v_0 \approx 520$ \kmpers,
$x_0=66.7$\hMpc{} for simulation BSI-500.  The minimum mass included in the
catalogs was $2.6\times 10^{11} \Msun$ for the high-resolution simulation
BSI-200*, $2\times 10^{11} \Msun$ for the simulations BSI-200 and CDM-200, and
$3 \times 10^{13} \Msun$ for the simulations BSI-500 and CDM-500.  However,
varying the minimum mass had little effect (i.e., resulted in a shift well
within the range of statistical uncertainty).  The Lauer and Postman (1994)
measurement $(689 \pm 178)$ \kmpers at 150\hMpc{} lies above the {\it bulk
velocity\/} in {\it any\/} of 512 bins for all CDM and BSI simulations at the
largest radius computed (75\hMpc).

An important property of BSI models as studied in GMS94 is to provide a natural
mechanism for a relatively high Mach-number or ``cold" flow (Ostriker \&{} Suto
1990, Strauss \ea{}, 1993).  Small-scale velocity fields may be studied by
measuring the rms line-of-sight relative peculiar velocity derived from galaxy
catalogs (Turner 1976; Davis \&{} Peebles 1983; Mo \ea{} 1993).

Davis and Peebles (1983) first inferred the small-scale velocity field of
galaxy
clustering from the anisotropy of the small-scale velocity dispersion.  They
derived the rms relative radial peculiar velocities $v_{\mbox{rms}}^2 =
\langle (v_1 - v_2)^2 \rangle$ of galaxy pairs (where $v_1$ and $v_2$ are
the radial peculiar velocities) as a function of the projected distance $r_p$
between
them on the sky. In the simulations we derive the relative peculiar streaming
motion
as function of the projected distance in bins up to 4 \hMpc{}, and the rms of
the
relative peculiar velocity differences for galaxies with projected separation
between 1 and 2 \hMpc, \cp{} Table 3.  While the streaming motion falls off
with increasing $r$, the velocity dispersion $v_{\mbox{rms}}$ is almost
constant,
therefore we give only one value.  To restrict the contamination due to
foreground and background galaxies, we consider only pairs of galaxies with
{\sl
real} separation less than 10 \hMpc{}.  The values of the Table 3 are directly
comparable to the Davis and Peebles (1983) results which give rms peculiar
velocities from the apparent anisotropy of the spatial correlation function
using
the CfA redshift survey:
$v_{\mbox{rms}}(R)=(340 \pm 40) (r_p \mbox{h}^{-1}\mbox{Mpc})^{0.13}$ \kmpers,
for $r < 1$\hMpc.  We get somewhat smaller velocity differences for all BSI
simulations, which result from the combined effect of restricted power in the
simulation box, and the insufficient resolution of small mass elements. In
particular, we cannot resolve single galaxies in the large simulation boxes,
but only infer the peculiar velocities of 'groups' or 'clusters'.  The highest
peculiar velocities are predicted for the 75 \hMpc{} and the high resolution
400 \hMpc{} simulations.  On the contrary, the velocity dispersion of CDM
galaxies is too high.

\begin{table*}
\begin{minipage}{120mm}
\caption{First ($v_{\mbox{str}}$, defined positive inwards) and  second
($v_{\mbox{rms}}$) moments of distribution of line-of-sight projected relative
peculiar velocity differences of galaxy halo pairs.}
\begin{tabular}{lcccccc}
   \hline
   Simulation & $v_{\mbox{str}}$ [{km~s${}^{-1}$}] & & & &
   & $v_{\mbox{rms}}$ [{km~s${}^{-1}$}] \\
   & (0 - 1) & (1 - 2) & (2 - 3) & (3 - 4)
   & & (1 - 2) \\
   & \hMpc & \hMpc & \hMpc & \hMpc & & \hMpc\\
   \hline
BSI-200*& 183 & 159 & 142 & 130 &   &  256 \\
BSI-25* & ~50 & ~40 & ~27 & ~18 &   &  192 \\
BSI-500 & 176 & 169 & 139 & 106 &   &  243 \\
BSI-200 & 141 & 117 & 111 & 101 &   &  241 \\
BSI-75  & 106 & 104 & ~97 & ~88 &   &  253 \\
BSI-25  & ~82 & ~78 & ~69 & ~62 &   &  192 \\
CDM-500 & 362 & 373 & 316 & 266 &   &  469 \\
CDM-200 & 224 & 216 & 201 & 172 &   &  484 \\
CDM-75  & 204 & 218 & 207 & 199 &   &  644 \\
CDM-25  & 100 & 126 & 131 & 110 &   &  469 \\
   \hline
\end{tabular}
\end{minipage}
\end{table*}

Recent studies of the same and newer data sample seem to indicate that the
small-scale velocity dispersion depends on the selection of galaxy types,
especially whether large clusters are included or not (Mo \ea{}, 1993).  In the
latter case, a significantly higher velocity dispersion is obtained.  Also
studying the groups and clusters in simulations, one gets higher rms
velocities.
These results do not refer to the analysis of the mean of relative velocities
of {\sl all} close projected pairs as done here.

\section{Conclusions}
\label{concl}

A substantial body of independent and plausible observational evidence
indicates
that the primordial perturbation spectrum is not well described by the standard
CDM model with Harrison-Zel'dovich primordial fluctuation spectrum.  Here, we
have seen that the two-parameter spectrum with broken scale invariance
resulting
from a double-inflationary scenario provides a good description to a
substantial
range of observational data:  the galaxy number density, galaxy-galaxy
correlations, the count-in-cell statistics of galaxies, the cluster mass and
X-ray temperature functions, cluster-cluster correlations, small and
large-scale
velocity fields.  The observed filamentary structure is evident in the galaxy
distribution, the distribution of cold particles and (most prominently) in the
distribution of warm ($\approx 10^6$~K) gas.

With regard to counts in cells, the very encouraging result --- obtained from
the distribution of DM particles -- is that the BSI simulations fit the data
quite well in a wide range of scales, [$(5 - 60)$\hMpc].  In contrast, the CDM
counts in cells are inconsistent with observations at about the 2-$\sigma$
level.

The variance of {\sl galaxy} counts depends strongly on the chosen galaxy
identification scheme.  The results are summarized in Table 2 with the
estimated
integral bias.  For simulations BSI-200 and BSI-200*, the results are at best
typical for galactic halos, while in BSI-500 we measure instead cluster halos.
In contrast, for the CDM simulations we found 'galaxies' that are slightly
anti-biased with respect to the dark-matter distribution.

BSI seems to produce fewer massive objects than CDM. However, in view of
overmerging, care must be taken in interpreting the apparent disagreement of
mass spectra derived in either scenario with mass spectra inferred from
luminosity functions.

The slope and correlation length of the BSI galaxy correlation function as
discussed
in Section 7 are close to observed values.  The cluster correlation function
has
a correlation radius of $(10 - 15)$\hMpc{} and stays positive at least out to
about
80\hMpc, consistent with our observational knowledge.

Rms line-of-sight velocities from BSI simulations appear to be closer to the
Davis \&{} Peebles (1983) data than comparable CDM simulations, which lie
systematically too high as mentioned already in the first papers on CDM
simulations.

Large-scale streaming velocities in the 500\hMpc{} simulations are well within
1-$\sigma$ statistical variations as determined from the simulations.
Considering the current uncertainties in these measurements, they pose no
problems for the model at this time.  Moreover, a circa 25 per cent increase in
the
normalization as suggested by the G\'orski \ea{} (1994) reanalysis of the COBE
data would suggest roughly a 7 per cent boost in typical large-scale motions at
the
scales in question. But in any case our predictions are much lower than the
Lauer \&{} Postman (1994) results, our BSI expectation at a scale of 150
\hMpc{}
is about 55 km/s.

 An important test for the BSI spectrum will be the more precise measurement of
multipole moments of the microwave anisotropy at scales of a few hundred Mpc
(the Doppler peak).  The advantage of these measurements is that they directly
sample the linear part of the spectrum.  A decisive question for BSI will be a
more precise observational determination of the epoch of formation of the first
objects.  As mentioned in Sect.~5, improved treatment of ionization and heating
processes could allow use of quasar abundance and quasar absorption line data
to
obtain more stringent limits on structure evolution in BSI.  Of comparable
importance are data expected to be available in the near future on the
evolution of the cluster mass function at redshifts $z\approx 0.5$.  This
problem
presents a theoretical challenge, because it requires an accurate
hydrodynamical
treatment of cluster gas dynamics in the context of large-scale structure
simulations.

The most important success of the BSI scenario is the excellent agreement with
present observational data on very large scale structure. In this paper, we
have seen through numerical simulations that smaller scales are generally in
excellent agreement with available data.  In the future, it will be important
to
investigate additional characteristics of large-scale structure using methods
such as
topological studies, percolation analysis and computation of void probability
functions.

In order to study mass spectra and statistics in a more reliable manner,
hydrodynamic simulations should be performed.  However, the results of our PM
simulations strongly suggest that the BSI model studied here deserves full
attention as a reliable description of cosmological structure formation.
\bigskip

\noindent {\bf Acknowledgements:}

We would like to express our thanks to Andrei Dorosh\-kevich and Anatoly Klypin
for stimulating discussions.  Our referee, David Weinberg, made a lot of
constructive
remarks which helped us very much in improving the paper.  Karl-Heinz B\"oning
provided invaluable support in computer management.


\begin{thebibliography}{99}

\bibitem{agmm95} Amendola, L.,  Gottl\"ober, S., M\"ucket, J. P., \&{}
M\"uller, V.
              1995, ApJ in press
\bibitem{b83} Bahcall, N. A., \&{} Soneira, R. A. 1983, ApJS, 70, 1
\bibitem{bc93} Bahcall, N. A., \&{} Cen, R. 1993, ApJ. 407, L49
\bibitem{bs93} Bartlett, J. G.,  \&{} Silk, J. 1993, ApJ. 407, L45
\bibitem{be93} Baugh, C. M., \&{} Efstathiou, G. 1993, MNRAS 265, 145
\bibitem{bdfdb90} Bertschinger, E., Dekel, A., Faber, S., Dressler, A., \&{}
Burstein, D.
                  1990 ApJ 364, 370
\bibitem{b82} Blumenthal, G. R., Pagels, H., \&{} Primack, J. R. 1982, Nature
299, 37
\bibitem{be84} Bond, J. R., \&{} Efsthathiou, G. 1984, ApJ 285, L45
\bibitem{b94} Bond, J. R., 1994, CITA-94-5 (Proceedings of the Capri meeting)
\bibitem{cgko92} Cen, R., Gnedin, N.Y., Kofman, L. A., \&{} Ostriker, J.  1992,
              ApJ 399, L11
\bibitem{cgo93} Cen, R., Gnedin, N.Y., \&{} Ostriker, J. 1993, ApJ 417, 387
\bibitem{co92a} Cen, R., \&{} Ostriker J. 1992a, ApJ 392, 22
\bibitem{co94} Cen, R., \&{} Ostriker J. 1994, ApJ 431 451
\bibitem{c94} Cen, R. 1004, ApJ 437, 12
\bibitem{co94} Collins, C. A., Cruddace, R. G., Ebling, H.,
   MacGillivray, H. T., \&{} Voges, W. 1994 in: 'Studying the Universe
   with clusters of galaxies',
   eds. H. B\"ohringer, S. Schindler, MPE Report 256, 107
\bibitem{dss92} Davis, M., Summers, F. J. \&{} Schlegel, D. 1992, Nature 359,
393
\bibitem{dp83} Davis, M. \&{} Peebles, P.J.E. 1983, ApJ 267, 465
\bibitem{dp83} Doroshkevich, A. G., Fong, D., Gottl\"ober, S., M\"ucket, J. P.,
\&{}
               M\"uller, V. 1995 MNRAS submitted
\bibitem{eep88} Efstathiou, G., Ellis, R.S., \&{}  Peterson, B.S. 1992,
                MNRAS 232, 431
\bibitem{e92} Efstathiou, G. 1992, MNRAS 256, 43P
\bibitem{ebw92} Efstathiou, G., Bond, J. R., \&{} White, S. 1992, MNRAS 258, 1P
\bibitem{esm90} Efstathiou, G., Kaiser, N., Saunders, W., Lawrence, A.,
         Rowan-Robinson, M., Ellis, R.S., \&{} Frenk, C. S. 1990a, MNRAS 247,
10P
\bibitem{e90} Evrard, A.E. 1990, ApJ. 363, 349
\bibitem{emfg94} Evrard, A.E., Mohr, J.J., Fabricant, D.G., \&{} Geller, M.J.
1993,
                Astr.J.  95, 985
\bibitem{fr85} Fall, S.M., \&{} Rees, M. 1985, ApJ 298, 18
\bibitem{fdsyh93} Fisher, K. B., Davis, M., Strauss, M. A. Yahil, A., \&{}
Huchra, J. P.
                  1993, ApJ 402, 44
\bibitem{g94} G\'orski, K. M., Hinshaw, G., Banday, A. J., Bennett, C. L.,
Wright, E. L., Kogut, A., Smoot, G. F., \&{} Lubin, P. 1994, ApJ 430, L89
\bibitem{g94} Gottl\"ober, S., \&{} M\"ucket, J. P. 1993, A$\&{} $A 272, 1
\bibitem{gms94} Gottl\"ober, S., M\"ucket, J. P., \&{} Starobinsky A. A. 1994,
                ApJ, 434, 417, GMS94
\bibitem{gms91} Gottl\"ober, S., M\"uller, V., \&{} Starobinsky, A. A. 1991,
                Phys. Rev. D43, 2510 GMS91
\bibitem{g94} Gottl\"ober, S. 1994, in: 'Studying the Universe with clusters of
              galaxies', Ringberg workshop, eds. H. B\"ohringer, S. Schindler,
              MPE Report 256, 79
\bibitem{ha91} Henry, J. P., \&{} Arnaud, K. A. 1991, ApJ 372, 410
\bibitem{k85} Kaiser, N. 1984, ApJ 284, L9
\bibitem{k91} Kaiser, N. 1991, ApJ 383, 104
\bibitem{kss94} Kamionkowski, M. Spergel, D., \&{} Sugiyama, N. 1994, ApJ 426,
L57
\bibitem{kocrjdbn94} Kang, H., Ostriker, J.P., Cen, R., Ryu, D., Hernquist, L.,
                     Evrard, A.E., Bryan, G.,\&{} Norman, M.L., 1994, ApJ 430,
83
\bibitem{kkk91} Kates, R., Kotok E., \&{} Klypin, A. 1991, A\&{} A 243, 295
(KKK91)
\bibitem{kg91} Katz N., \&{} Gunn, J. 1991, ApJ 377, 365
\bibitem{khw92} Katz N., Hernquist, L. \&{} Weinberg, D. H. 1992, ApJ 399, L109
\bibitem{kw93} Katz N., \&{} White, S. 1993, ApJ. 412, 455
\bibitem{kwg93} Kauffmann, G., White, S., \&{} Guiderdoni, B. 1993, MNRAS 264,
201
\bibitem{kc94} Kauffmann, G., \&{} Charlot, S. 1994, ApJ 430, L97
\bibitem{khhpr91} Klypin, A., Holtzman, J. Primack. J., \&{} Reg\"os, E. 1993,
                  ApJ 416, 1
\bibitem{kk91} Klypin, A., \&{} Kates, R. 1991, MNRAS 251, 41P (KK)
\bibitem{kkk92} Klypin, A., Kates, R, \&{} Khokhlov, A. 1992, in:
               'New Insights into the Universe', Lecture Notes in Physics 171
                Springer, 157 (KKK92)
\bibitem{kk83} Klypin, A., \&{} Kopylov, A. I. 1983, Soviet Astr. Letters 9, 41
\bibitem{kgb93} Kofman, L., Gnedin, N., \&{}  Bahcall, N. 1993, ApJ 413, 1.
\bibitem{lp94} Lauer, T., \&{} Postman, M. 1994, ApJ 425, 418.
\bibitem{l92} Lilje, P. B. 1992, ApJ 386, L33
\bibitem{lpem92} Loveday, J., Peterson P., Efstathiou, G., \&{} Maddox, S.
1992,
                 ApJ 390, 338
\bibitem{mesl90} Maddox, S. J., Efstathiou, G., Sutherland, W.J., \&{} Loveday,
J.
                 1990, MNRAS  242, 43
\bibitem{m93} Mo, H.J., Jing, Y.P., B\"orner, G. 1993, in: B\"orner \&{}
Buchert
              (eds.), `Proc. 4. MPG-CAS Workshop on High Energy Physics and
              Cosmology' MPA/P8
\bibitem{mm94} Mo, H.J.,  \&{} Miralda-Escud\'{e}, J. 1994, ApJ 430, L25
\bibitem{mkpr95} M\"ucket, J.P., Kates, R., Petitjean, P., \&{}  Riediger, R.
1995,
                 in preparation
\bibitem{m94a} M\"uller, V. 1994a, in: 'Cosmological Aspects of X-Ray Clusters
of
              Galaxies', NATO ASI Series Vol 441, Kluwer, 439
\bibitem{m94b} M\"uller, V. 1994, in: 'Studying the Universe with clusters of
              galaxies', Ringberg workshop, eds. H. B\"ohringer, S. Schindler,
              MPE Report 256, 85
\bibitem{nb91} Navarro, J. F., \&{} Benz, W. 1991, ApJ 380, 320
\bibitem{o93} Ostriker, J. 1993, Ann. Rev. Astron. Astrophys. 31, 689
\bibitem{os90} Ostriker, J., \&{} Suto 1990, ApJ 348, 378
\bibitem{p92} Park, C., Gott, J., \&{} da Costa, L. 1992, ApJ 392, L51
\bibitem{p95} Petitjean, P., M\"ucket, J. P., \&{} Kates, R. 1995, A$\&$A 295,
L9
\bibitem{ps74} Press, W. H., \&{} Schechter, P. 1974, ApJ 187, 425
\bibitem{s} Shapiro, P., \&{} Struck-Marcell, C. 1985, ApJ Suppl 57, 205
\bibitem{s92} Smoot, G. F. et al. 1992, ApJ 396, 1
\bibitem{s85} Starobinsky, A. A., 1985, JETP Lettt. 42, 152
\bibitem{sm94} Steinmetz, M., \&{} M\"uller, E. 1995, MNRAS in press
\bibitem{tw94} Thoul, A.A., \&{} Weinberg, D.H. 1994, preprint astro-ph/941000
\bibitem{t76} Turner, E. L. 1976, ApJ 208, 20
\bibitem{vpgh92} Vogeley, M. S., Park, C., Geller, M., \&{} Huchra, J. P. 1992,
                 ApJ 391, L5
\bibitem{wfde87} White, S., Frenk, C. S., Davis, M., \&{} Efstathiou, S. 1987,
                 ApJ 313, 505
                 MNRAS 258, 1P
\bibitem{wnef94} White, S., Navarro, J. F., Evrard, A. E., \&{} Frenk, C. S.
1994,
                 Nature 366, 429
\bibitem{z70} Zel`dovich, Y. B. 1970, A$\&$A 5, 84
\end{thebibliography}
\end{document}